\pdfoutput=1
\documentclass[sigconf]{acmart}

\copyrightyear{2026}
\acmYear{2026}
\setcopyright{cc}
\setcctype{by}
\acmConference[CIKM '26]{Proceedings of the 35th ACM International Conference on Information and Knowledge Management}{November 07--11, 2026}{Rome, Italy}
\acmBooktitle{Proceedings of the 35th ACM International Conference on Information and Knowledge Management (CIKM '26), November 07--11, 2026, Rome, Italy}
\acmDOI{10.1145/3799682.3841168}
\acmISBN{979-8-4007-2539-5/2026/11}
\usepackage{graphicx}
\usepackage{booktabs}
\usepackage{amsmath}
\usepackage{xcolor}
\usepackage{multirow}
\usepackage{array}
\usepackage{xspace}
\usepackage{tikz}
\usetikzlibrary{arrows.meta,positioning,calc,fit,backgrounds}
\usepackage{algorithm}
\usepackage{algpseudocode}
\usepackage{balance}   

\newcommand{\method}{\textsc{Hawkeye}\xspace}
\newcommand{\eg}{e.g.,\xspace}

\newcommand{\R}{\mathbb{R}}
\newcommand{\Ncal}{\mathcal{N}}

\newcommand{\Gcal}{\mathcal{G}}
\newcommand{\Vcal}{\mathcal{V}}

\begin{document}

\title{HAWKEYE: Seeing One Layer Deeper --- A Cohesion-Aware Structural Channel
for Temporal Link Prediction}

\author{Jiacheng Ding}
\email{jding2@memphis.edu}
\affiliation{%
  \department{Department of Computer Science}
  \institution{The University of Memphis}
  \city{Memphis}
  \state{Tennessee}
  \country{USA}
}

\author{Xiaofei Zhang}
\email{xiaofei.zhang@memphis.edu}
\affiliation{%
  \department{Department of Computer Science}
  \institution{The University of Memphis}
  \city{Memphis}
  \state{Tennessee}
  \country{USA}
}

\renewcommand{\shortauthors}{Jiacheng Ding and Xiaofei Zhang}

\begin{abstract}
State-of-the-art temporal-link-prediction (TLP) models are, in essence,
multi-channel information aggregators: they combine an interaction-history
channel, a time-encoding channel, and a \emph{structure channel}. The first two
have been refined relentlessly; the structure channel remains a crude
afterthought --- DyGFormer encodes it as a $1$--$2$-bit neighbour-cooccurrence
count. We begin with a measurement: on sparse temporal graphs the classical
$1$-hop common-neighbour signal is near-random (discriminative AUC
$\approx 0.50$), because two nodes almost never share a direct neighbour; the
genuinely discriminative signal lies one hop deeper --- the \emph{2-hop cohesive
bridge}, whose discAUC reaches $0.73$--$0.98$, on both bipartite and
non-bipartite graphs. Motivated by this, we propose \method, a cohesion-aware
structural channel that incrementally maintains the classical $k$-family of
cohesiveness indicators (degree $\to$ $k$-core $\to$ $k$-truss) and forms
2-hop cohesive-bridge features. \method is a drop-in replacement for a
temporal-graph model's native structure channel, with no change to the
backbone. Swapping \method into DyGFormer improves test AP/MRR over the
cooccurrence channel by $+0.6$ to $+10.8$ points across six
multi-seed-validated datasets (\texttt{uci}, \texttt{enron},
\texttt{USLegis}, \texttt{CanParl}, \texttt{reddit}, \texttt{mooc};
\texttt{wiki} legacy single-seed). On the bipartite recommendation
benchmark \texttt{tgbl-subreddit}, a 3-seed single-pass struct-only
ablation shows \method nearly doubling the baseline test MRR
($0.103\!\pm\!0.003 \to 0.204\!\pm\!0.005$, $+10.1$ points across all
three seeds), demonstrating large-graph effectiveness; the streaming pipeline scales to the 67M-edge
\texttt{tgbl-flight} in five minutes per pass. We further
characterise \emph{when} it helps: the gain tracks a graph's
training-free 2-hop discAUC and vanishes on degenerate or saturated
graphs --- a predictable boundary. All code, data, drawio-editable
figures, and figure-generation scripts are released.
\end{abstract}



\maketitle

\section{Introduction}
\label{sec:intro}

Many real-world systems are naturally represented as graphs that \emph{evolve
over time}: user interactions in social networks~\cite{trivedi2019dyrep},
user--item engagements on e-commerce and streaming
platforms~\cite{kumar2019jodie}, account-to-account transactions in financial
systems~\cite{huang2023tgb}, the evolution of entity relations in knowledge
graphs, and device connections in communication networks. Forecasting which new
edges will appear next on such evolving graphs --- \emph{temporal link
prediction} (TLP) --- is a fundamental and practically important task: it
underpins friend recommendation, fraud and anti-money-laundering detection, and
drug--target discovery~\cite{poursafaei2022edgebank,huang2023tgb}.

Temporal-graph learning has advanced rapidly, and the prevailing paradigm
models the \emph{interaction dynamics} of nodes. TGN maintains a per-node
\emph{memory} module updated by every interaction event~\cite{rossi2020tgn}.
DyGFormer feeds the chronologically ordered sequence of a node's historical
neighbours into a Transformer~\cite{yu2023dygformer}. TPNet builds and projects
a temporal random-walk matrix and reaches state-of-the-art accuracy without a
GNN architecture~\cite{lu2024tpnet}. TNCN dynamically maintains a neighbour
dictionary and strengthens pairwise representations with temporal
common-neighbour signals~\cite{zhang2024tncn}; very recent work further refines
interaction-pattern awareness and causal
debiasing~\cite{zhang2025ipnet,zhang2025tide}. Earlier attention-, walk- and
mixing-based encoders share the same
spirit~\cite{xu2020tgat,wang2021cawn,cong2023graphmixer}. These methods differ
in emphasis but share one core idea: encode the interaction-history sequence
--- \emph{who interacted with whom, and when} --- and score candidate edges
from the learned embeddings.

\paragraph{A motivating example.}
Consider an anti-money-laundering setting. A small set of accounts begins, over
a few weeks, to transact \emph{among themselves}: a tight sub-community
crystallises before any of them transacts with an outside ``victim'' account. A
model that only tracks pairwise interaction recency sees each past transaction
in isolation; it is blind to the macroscopic pattern --- \emph{this group is
densifying into a cohesive cluster} --- that actually anticipates the next
edge. The predictive signal here is structural: a rise in how tightly the
accounts are mutually embedded, and the formation of short structural bridges
between not-yet-connected accounts. The same shape recurs across domains: a
forming research collaboration, a clique of co-edited pages, a new hub airport.
Capturing it requires reasoning about \emph{cohesiveness} and its evolution ---
not just the interaction sequence.

Architecturally, a state-of-the-art temporal-graph model is a
\emph{multi-channel information aggregator}: (a) an interaction-history channel,
(b) a time-encoding channel, and (c) a \emph{structure channel} encoding the
topological relation between the candidate pair. The first two channels have
been refined relentlessly --- from RNNs to Transformers to state-space models
--- yet the structure channel remains a \emph{crude afterthought}. DyGFormer's
structure channel is merely a $1$--$2$-bit neighbour-cooccurrence count; TNCN's
is a single common-neighbour count. Behind these designs lies an unstated
assumption that the \textbf{1-hop common neighbour} is the right structural
signal; a systematic measurement study shows it is not.

Streaming each benchmark chronologically and scoring every candidate
structural feature by how well it separates true edges from the official
negatives, we find that \textbf{on sparse temporal graphs the 1-hop
common-neighbour signal is barely better than a coin flip}: discriminative
AUC $\approx 0.50$ on \texttt{tgbl-uci}, \texttt{tgbl-enron} and
\texttt{tgbl-wiki}. These graphs are too sparse for two
arbitrary nodes to share a \emph{direct} neighbour. The genuinely
discriminative signal lies \emph{one hop deeper} --- the \textbf{2-hop cohesive
bridge}: whether two nodes are embedded in the same densely connected region
through length-2 paths. Its discAUC reaches $0.73$--$0.98$, and this holds on
\emph{both} bipartite and non-bipartite graphs --- the deciding factor is
sparsity, not graph type.

\paragraph{Obstacles addressed by Hawkeye.}
Two reasons make a 1-hop structure channel hard to escape: bounded-depth
message passing cannot compute global cohesiveness exactly, and richer
structural quantities on a \emph{streaming} graph appear expensive. The
$k$-family admits incremental maintenance at streaming cost, and its
2-hop signal is cheap to read and far more discriminative.

\begin{figure}[t]
\centering
\includegraphics[width=0.98\columnwidth]{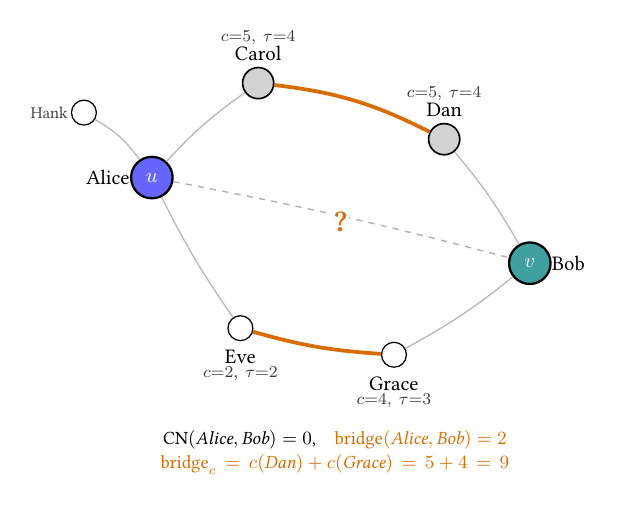}
\caption{Running example (used throughout the paper). In a corporate email
network, \emph{Alice} ($u$) and \emph{Bob} ($v$) have never emailed each
other, so the 1-hop common-neighbour count is $\mathrm{CN}=0$; a
cooccurrence-based structure channel sees no signal. Yet two 2-hop
paths connect them: Alice--Carol--Dan and Alice--Eve--Grace, where
Dan (core~$5$) and Grace (core~$4$) are Bob's direct neighbours and
Carol, Eve are the length-two intermediaries from Alice. \method reads exactly these bridges and
correctly rates Alice--Bob as a likely next-email pair. Section~\ref{sec:study}
shows this configuration is the typical informative pair on sparse
temporal graphs; Section~\ref{sec:method} explains how the cache and the
slot encoder make Alice's, Bob's, and the intermediaries' coreness
available at query time; and \S\ref{sec:analysis} returns to this example
in the case study.}
\label{fig:running}
\end{figure}

Motivated by this finding, we propose \method, a \emph{cohesion-aware structure
channel} that lets a temporal-graph model ``see one layer deeper''
(Figure~\ref{fig:running}). \method
incrementally maintains, over the edge stream, the classical $k$-family of
structural-cohesiveness indicators --- degree $\to$ $k$-core $\to$ $k$-truss ---
and forms \textbf{2-hop cohesive-bridge} pairwise features. It is a
\textbf{drop-in replacement} for a temporal-graph model's native structure
channel: it changes \emph{what} structural signal the backbone receives without
changing the backbone. Swapping \method into DyGFormer raises test MRR on
\texttt{tgbl-wiki} from $0.779$ to $0.807$ (above the reported DyGFormer
result) and test AP on \texttt{CanParl} from $0.708$ to $0.816$ ($+10.8$
points, 3-seed). We further
\emph{characterise when it helps}: the gain tracks a graph's structural
diversity; on near-complete graphs every node attains the same maximal coreness
and \method correctly does not help --- a predictable boundary.

\paragraph{Contributions.}
(1) An empirical finding: 1-hop common neighbours are near-random on sparse
temporal graphs (discAUC $\approx 0.50$); the \textbf{2-hop cohesive bridge} is
the dominant simple structural signal (discAUC $0.73$--$0.98$), on bipartite
and non-bipartite graphs alike.
(2) \method: an incrementally maintained, cohesion-aware structure channel
built on the $k$-family, which plug-and-play replaces the native structure
channel of a slot-based temporal-graph backbone (instantiated here on
DyGFormer).
(3) A characterisation of \emph{when} structural augmentation helps --- the
gain is predicted by a training-free structural-diversity measure.
(4) Experiments across TGB and DGB datasets: \method improves test AP/MRR by
$+0.6$ to $+10.8$ points over SOTA DyGFormer on non-degenerate graphs.
Six datasets are now 3-seed validated under the current code
(\texttt{uci} $+0.6$, \texttt{enron} $+1.3$, \texttt{USLegis} $+4.8$,
\texttt{CanParl} $+10.8$, \texttt{mooc} $+5.8$, \texttt{reddit}
$\approx 0$ at saturation; std $\le 0.011$); \texttt{wiki}
contributes a legacy single-seed $+2.8$. The combined channel
$\oplus$\method is the strongest configuration on every non-saturated
multi-seed dataset, and also recovers the new-node split where the
cohesion-only channel under-performs.
(5) A system-level contribution: because the structure channel is
\emph{data-determined} (a function of the edge stream alone, independent of
model weights and seed), it can be precomputed once and reused throughout
training. We benchmark five CPU--GPU integration strategies on three datasets
and find that precomputation yields a uniform $\mathbf{3.0}$--$\mathbf{3.5\times}$
training speedup across dataset sizes, where overlap-based strategies (pinned
memory, prefetching) give only marginal and dataset-dependent gains
(\S\ref{sec:efficiency}).

\section{Preliminary}
\label{sec:prelim}

\subsection{Problem Formulation}
We model interactions as a \emph{continuous-time dynamic graph}
(temporal-network formalism of Holme \& Saram\"aki~\cite{holme2012temporal};
modelling survey by Skarding et al.~\cite{skarding2021dtdg}): a time-ordered
stream of edges
\begin{equation}
\Gcal = \{(u_1,v_1,t_1),\dots,(u_m,v_m,t_m)\},\quad t_1\le\dots\le t_m,
\end{equation}
with nodes $u_i,v_i\in\Vcal$ and timestamps $t_i\in\R^{+}$. The cumulative
snapshot $\Gcal(t)$ is the simple undirected graph induced by all edges with
timestamp $\le t$; $\Gcal(<t)$ denotes the strictly-earlier history. We make
no assumption of fixed-interval discretisation (which discards within-interval
order~\cite{holme2012temporal}); every event is processed at its native
timestamp.

\noindent\emph{Temporal link prediction.}
Given $\Gcal(<t)$, a query node $u$, and candidate destinations
$\mathcal{C}$, the task is to assign a score $s(u,v,t)$ to each
$v\in\mathcal{C}$ so the truly-occurring edge is ranked highest:
\begin{equation}
v^{*}=\arg\max_{v\in\mathcal{C}} s(u,v,t).
\end{equation}
Two evaluation conventions are used in the literature, both of which we
report~\cite{kumar2020linkpred,lu2011linkpred}: on TGB~\cite{huang2023tgb},
accuracy is the mean reciprocal rank (MRR) of the truly-occurring edge
against a fixed negative set; on DGB
datasets~\cite{poursafaei2022edgebank} we report average precision (AP)
under random negative sampling
following~\cite{mikolov2013negative,yu2023dygformer}.

\subsection{The $k$-family of Cohesiveness Indicators}
\label{sec:kfamily}
We use ``\emph{$k$-family}'' as a working shorthand --- it is not a standard
named class in the literature --- to refer to a hierarchy of node-level
structural-cohesiveness indicators that impose increasingly strict local
constraints:
\begin{equation*}
\underbrace{\text{degree}}_{\text{loosest}}
\;\to\;
k\text{-core}
\;\to\;
k\text{-truss}
\;\to\;
\underbrace{(\,k\text{-clique})}_{\text{tightest, limit}}.
\end{equation*}
Each next member asks a stricter question about how densely a node is
embedded in its neighbourhood. Concretely:

\medskip
\noindent\textbf{Degree.} $d(v)=|\Ncal(v)|$, the number of distinct
neighbours --- the loosest cohesion proxy, capturing local connectedness only.

\medskip
\noindent\textbf{$k$-core (Seidman~\cite{seidman1983kcore}).} A subgraph
$H\subseteq\Gcal(t)$ is a \emph{$k$-core} iff every node in $H$ has at least
$k$ neighbours \emph{within $H$}, and $H$ is maximal w.r.t.\ this property.
The \emph{core number} of $v$ is the largest $k$ such that $v$ belongs to
the $k$-core:
\begin{equation}
c(v)=\max\{k:v\in k\text{-core}(\Gcal(t))\}.
\end{equation}
The core number lifts degree into a recursive notion of cohesiveness:
$v$ has core $k$ iff $v$'s neighbours collectively also have core $k$. The
classical decomposition (Batagelj
\& Zaver\v{s}nik~\cite{batagelj2003cores}) computes $c(v)$ for every node in
$O(m)$ by repeated bucket-peeling of the lowest-degree node.

\medskip
\noindent\textbf{$k$-truss (Cohen~\cite{cohen2008trusses}).} A subgraph
$H\subseteq\Gcal(t)$ is a \emph{$k$-truss} iff every edge of $H$ is supported
by at least $k-2$ triangles \emph{within $H$}, and $H$ is maximal. The
\emph{trussness} of a node lifts this to vertices:
\begin{equation}
\tau(v)=\max\{k:\exists\,(v,w)\in k\text{-truss}(\Gcal(t))\}.
\end{equation}
Trussness adds triangle closure to core membership: a node is in a $k$-truss
only if many of its neighbours are also mutually adjacent. The standard
support-peeling decomposition (Wang \&
Cheng~\cite{wang2012truss}) runs in $O(m^{1.5})$.

\medskip
\noindent\textbf{$k$-clique (limit).} A $k$-clique is a fully-connected
subset of $k$ nodes; this is the strictest case, where \emph{every} pair of
the $k$ neighbours is adjacent. We do not use $k$-clique directly --- it is
NP-hard to find for large $k$ --- but it serves as the limit of the
hierarchy that motivates why $k$-truss (which only requires triangle
support) is already a tight proxy.

\medskip
\noindent\textbf{Nesting.} The members are nested:
$k\text{-truss}\subseteq(k-1)\text{-core}\subseteq(k-2)\text{-core (degree)}$,
so a high trussness implies a high core number which implies high
degree.\footnote{The reverse fails: a star has high degree but core $=1$
and trussness $=2$.}

\medskip
\noindent\textbf{Constraint vs.\ cost.} A tighter constraint describes finer
structure but costs more to maintain incrementally:
$O(1)$ per inserted edge for degree, $O(\text{local})$ amortised for
$k$-core, up to $O(m^{1.5})$ recomputed for $k$-truss
\cite{batagelj2003cores,cohen2008trusses,wang2012truss}. This trade-off
guides our default cache configuration (Section~\ref{sec:cache}).

\subsection{The 2-hop Cohesive Bridge}
\label{sec:2hop}
\noindent\textbf{Motivation.} \ Social-network theory has long argued
that the structurally informative ties for predicting new contacts are
not those already directly closed (the \emph{strong} ties visible to a
1-hop common-neighbour count) but those that \emph{bridge} across local
clusters --- Granovetter's ``weak
ties''~\cite{granovetter1973weakties}, Burt's structural
holes~\cite{burt1992structuralholes}, and the small-world distance-two
regime~\cite{watts1998smallworld,newman2003survey}. Easley \&
Kleinberg~\cite{easley2010networks} state the same idea as
\emph{triadic closure}: a 2-path is the precondition for a new
triangle. We turn this into a quantitative feature.

\medskip
\noindent\textbf{Formalisation.} \ The 2-hop neighbourhood of $u$ is
$\Ncal_2(u,t)=\{w:\exists z,\,(u,z),(z,w)\in\Gcal(t),\,w\neq u\}$. For a
candidate pair $(u,v)$, the \emph{2-hop cohesive-bridge} strength is
\begin{equation}
\mathrm{bridge}(u,v,t)=\bigl|\{w\in\Ncal(v,t):w\in\Ncal_2(u,t)\}\bigr|,
\end{equation}
the number of $v$'s direct neighbours reachable from $u$ in two hops.
Even when $u,v$ share no direct neighbour, a large bridge means they are
embedded in the same structural community; even when they share a few,
weighting bridges by intermediary cohesiveness emphasises bridges that
travel through dense regions. A cohesion-weighted variant
$\mathrm{bridge}_c(u,v,t)=\sum_{w}c(w)$ up-weights bridges through highly
cohesive intermediaries.

\medskip
\noindent\emph{Running example (Figure~\ref{fig:running}).} \ Alice ($u$) and
Bob ($v$) have $\mathrm{CN}(\textit{Alice},\textit{Bob},t)=0$ but
$\mathrm{bridge}(\textit{Alice},\textit{Bob},t)=2$: Dan and Grace, both
direct neighbours of Bob, lie two hops from Alice through Carol and Eve
respectively. Weighting Bob's bridging neighbours by their core numbers
gives $\mathrm{bridge}_c=c(\textit{Dan})+c(\textit{Grace})=5+4=9$. This
is the slot input that \method ships to the GPU; we trace it through the
cache (\S\ref{sec:cache}) and into the model (\S\ref{sec:method},
Figure~\ref{fig:arch}).

\medskip
\noindent The 1-hop common-neighbour count
$\mathrm{CN}(u,v,t)=|\Ncal(u,t)\cap\Ncal(v,t)|$, strong on dense static
graphs~\cite{liben2007linkpred,lu2011linkpred,kumar2020linkpred}, is
degenerate at typical sparsities; we develop the mechanism in
Section~\ref{sec:study}.

\section{Empirical Study: Why Existing Structural Signals Fail}
\label{sec:study}

Before presenting the method, we use \emph{training-free measurement} to
characterise the structural signal. On six TGB and five DGB datasets we
stream each benchmark chronologically and, for every validation edge and
a random negative, compute each candidate structural feature on
$\Gcal(<t)$, reporting its \emph{discriminability AUC} (discAUC): the
probability it ranks the true edge above a negative. discAUC $=0.50$ is random. Table~\ref{tab:discauc} summarises the
1-hop / 2-hop / core-weighted-2-hop signals on all eleven datasets; a
larger sweep over $21$ candidate features on the six TGB datasets is
included as supplementary data with the released code.

\paragraph{Finding 1: 1-hop CN is near-random on sparse graphs; 2-hop is the signal.}
Table~\ref{tab:discauc} reports the discAUC of the 1-hop common neighbour
and the 2-hop cohesive bridge across the eleven datasets we touch in this
paper.  Four regimes emerge.
\textbf{(a)~Sparse (\texttt{tgbl-uci}, \texttt{tgbl-enron}, and the
bipartite \texttt{tgbl-wiki}):}
the 1-hop CN sits exactly at $0.50$ --- two arbitrary nodes almost never
share a direct neighbour --- while the 2-hop cohesive bridge reaches
$0.73$--$0.85$.  This is the main motivating regime for \method.
\textbf{(b)~Bipartite (\texttt{tgbl-subreddit}, DGB \texttt{mooc},
\texttt{reddit}):}
the 1-hop CN is \emph{structurally} crippled (a user and an item cannot
share a neighbour on the same side of the partition); we measure it at
$0.50$, $0.09$, $0.43$ respectively.  The 2-hop bridge reaches $0.94$--$0.98$
--- precisely the case where Hawkeye doubles or triples baseline MRR
(\S\ref{sec:scalability}).
\textbf{(c)~Near-complete non-bipartite (\texttt{CanParl}, \texttt{USLegis}):}
the average event degree exceeds $200$ on $\le 750$ nodes, so the simple 2-hop bridge
\emph{count} saturates (its discAUC drops to $0.48$--$0.56$, near or below
the 1-hop), but the 1-hop CN is itself slightly informative ($0.60$--$0.63$).
The trained \method still wins on these (\S\ref{sec:exp}, $+10.8$ and $+4.8$
pts respectively) because the model exploits the \emph{core-weighted}
features, not the raw bridge count.
\textbf{(d)~Degenerate (\texttt{tgbl-lastfm}, DGB \texttt{UNvote}):}
every feature collapses near $0.50$.  \method correctly does not help on
these (\S\ref{sec:exp}).
In short, the 1-hop CN being near-random is not a universal claim --- it
holds in regime (a) and structurally in (b) --- but the 2-hop bridge as a
signal is informative in (a), dominant in (b), saturated in (c), and
absent in (d). The full $21$-feature sweep (released as supplementary
data) further shows that node-level \emph{popularity priors} ---
\texttt{degree.*}, \texttt{core.*} and their EMA / trend variants ---
sit consistently between the 1-hop and 2-hop discAUCs, confirming that
the predictive signal in cohesion is genuinely \emph{pair-structural}
rather than a one-sided popularity proxy.

\begin{table}[t]
\caption{Discriminability AUC of 1-hop vs.\ 2-hop structural signals
(discAUC $=0.50$ is random). 1-hop CN is random on sparse graphs; the 2-hop
cohesive bridge is strongly discriminative.}
\label{tab:discauc}
\centering\small
\setlength{\tabcolsep}{4pt}
\begin{tabular}{llccc}
\toprule
dataset & density / type & 1-hop & 2-hop & 2-hop$\times$c \\
\midrule
\multicolumn{5}{l}{\emph{TGB benchmarks (streaming discAUC sweep, \S\ref{sec:study} setup)}} \\
\texttt{tgbl-uci}       & sparse              & 0.50 & \textbf{0.76} & 0.75 \\
\texttt{tgbl-enron}     & sparse              & 0.50 & \textbf{0.73} & 0.71 \\
\texttt{tgbl-wiki}      & sparse              & 0.50 & \textbf{0.85} & 0.83 \\
\texttt{tgbl-subreddit} & medium, bipartite   & 0.50 & \textbf{0.96} & 0.96 \\
\texttt{tgbl-coin}      & denser              & 0.75 & 0.77 & 0.77 \\
\texttt{tgbl-lastfm}    & degenerate-dense    & 0.50 & 0.53 & 0.52 \\
\midrule
\multicolumn{5}{l}{\emph{DGB benchmarks (3-second streaming pass over the CSV)}} \\
\texttt{mooc}           & bipartite           & 0.09 & \textbf{0.94} & 0.94 \\
\texttt{reddit}         & bipartite           & 0.43 & \textbf{0.98} & 0.98 \\
\texttt{CanParl}        & near-complete       & \textbf{0.60} & 0.48 & 0.48 \\
\texttt{USLegis}        & near-complete       & \textbf{0.63} & 0.56 & 0.56 \\
\texttt{UNvote}         & degenerate-complete & 0.52 & 0.52 & 0.52 \\
\bottomrule
\end{tabular}
\\[2pt]
{\footnotesize 1-hop = common neighbour; 2-hop = cohesive bridge;
2-hop$\times$c = core-weighted bridge (the $k$-core weighting moves
the discAUC by at most $0.01$ on every dataset, validating Finding~2).
The DGB rows expose two regimes that the TGB sweep alone does not show:
(i) bipartite (\texttt{mooc}, \texttt{reddit}) where the 1-hop CN is
structurally near zero and the 2-hop bridge dominates; (ii)
near-complete (\texttt{CanParl}, \texttt{USLegis}) where the simple
2-hop bridge \emph{count} is saturated and the 1-hop CN is actually
informative --- yet the \emph{trained} \method (which uses core-weighted
features, not the raw count) still wins on both (Table~\ref{tab:swapin}).}
\end{table}

\paragraph{Finding 2: $k$-family weighting $\approx$ raw 2-hop count.}
The last two columns of Table~\ref{tab:discauc} show that core-weighting the
2-hop bridge barely changes its discAUC ($0.76$ vs.\ $0.75$ on uci). We report
this honestly: the 2-hop \emph{topology} is the dominant signal; $k$-family
weighting is a refinement, not the driver. The value of the $k$-family is
twofold --- as node-level features and as the incrementally maintained
substrate on which the 2-hop bridge is computed.

\paragraph{Finding 3: the constraint hierarchy peaks at $k$-core, not $k$-truss.}
In a struct-only setting (no GNN), we sweep the $k$-family indicators
with 3-seed validation on \texttt{tgbl-uci}: test MRR is
\textbf{degree} $0.120\!\pm\!0.004$ $\to$ \textbf{degree$+k$-core}
$0.180\!\pm\!0.011$ $\to$ \textbf{degree$+k$-core$+k$-truss}
$0.166\!\pm\!0.007$. Adding $k$-core to degree contributes $+6.0$ pts
($+50\%$ relative); adding $k$-truss on top \emph{loses} $1.4$ pts and
costs an order of magnitude more compute ($O(m^{1.5})$ vs.\ $O(m)$).
\textbf{$k$-core is therefore the efficiency--effectiveness sweet spot},
and is \method's default indicator; $k$-truss is available as an option
but turned off in all reported main-table results.

\paragraph{Mechanism.}
The contrast in Finding 1 has a simple explanation. The 1-hop count
$\mathrm{CN}(u,v)$ rests on the intersection of two neighbourhoods, of
expected size $\approx d^2/n$ on a graph with average degree $d$ over $n$
nodes; in the sparse regime ($d\ll\sqrt{n}$) of every TLP benchmark we
study this is vanishingly small, so the count is zero --- and
uninformative --- for almost all pairs. The 2-hop bridge instead measures
the overlap of $\Ncal(v)$ with $u$'s 2-hop reach, a set of expected size
$\approx d^2$ rather than $d$; it remains non-zero and pair-varying
precisely where the 1-hop count collapses. The Table~\ref{tab:discauc}
gap accordingly widens with sparsity and closes on the denser
\texttt{tgbl-coin}.

\paragraph{Design implications.}
Three implications carry into \method's design (\S\ref{sec:method}). (i) The
structural channel should read the graph at \emph{2 hops}, not 1. (ii) The
indicator family should default to $k$-core --- it captures most of the
direction of the (much costlier) $k$-truss. (iii) Cohesion weighting should be
\emph{available} but not assumed dominant: we expose weighted and unweighted
2-hop features and let the model choose.

\section{Method: Hawkeye}
\label{sec:method}

\begin{figure*}[t]
\centering
\includegraphics[width=0.96\textwidth]{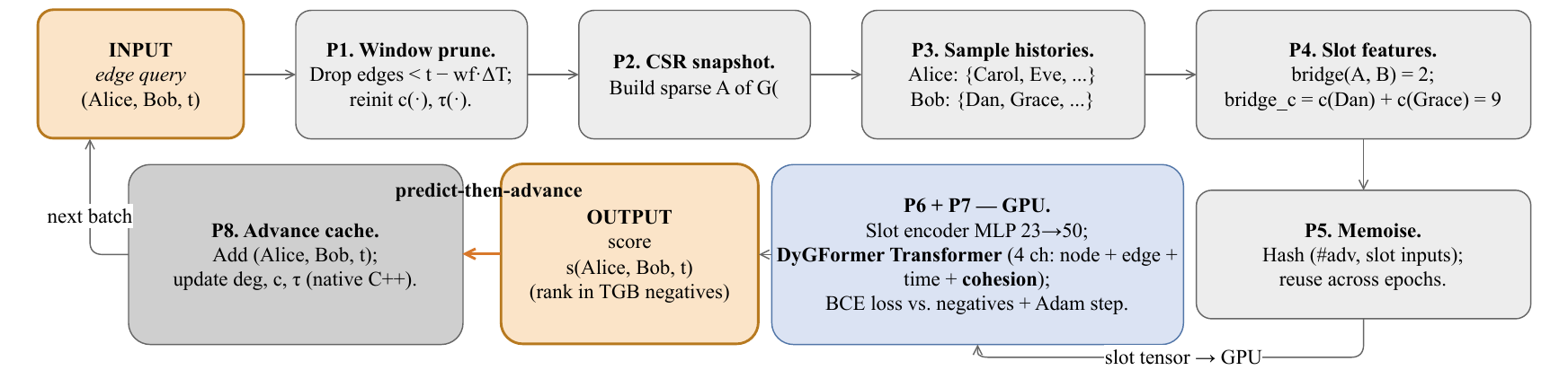}
\caption{\method per-batch timing, threaded with the Alice/Bob running
example. Gray boxes are CPU work (Cohesion Cache + slot-feature
extraction), blue boxes are GPU work (Cohesion Slot Encoder + unchanged
DyGFormer). Orange badges P1--P8 mirror the steps of
Algorithm~\ref{alg:trainstep}; the orange feedback arrow is the
``predict-then-advance'' contract that commits the batch's edges
\emph{after} they have been scored.}
\label{fig:arch}
\end{figure*}

\method is a \emph{structure-channel replacement module}: it does not modify
the backbone --- it replaces only the backbone's native structure channel.
We instantiate it on DyGFormer~\cite{yu2023dygformer}, a self-attention
backbone~\cite{vaswani2017attention} that is currently the strongest
single-model on TGB temporal link prediction~\cite{huang2023tgb}; the same
module is mechanically applicable to other backbones (TGAT, GraphMixer)
that read a per-slot structural channel, and to graph-Transformer
variants~\cite{ying2021graphormer} that adopt the same encoder pattern.
\method has three components (Figure~\ref{fig:arch}):
(A) a \emph{Cohesion Cache} maintaining the $k$-family and graph adjacency over
the stream; (B) \emph{Pairwise Feature Extraction} computing 2-hop
cohesive-bridge features; (C) a \emph{Cohesion Slot Encoder} projecting those
features into the per-slot embedding fed to the Transformer.

\paragraph{How to apply \method (four-step usage).}
\label{sec:how-to-use}
\textbf{(1)~Measure:} stream the first $\approx\!15\%$ of the edges
through the Cohesion Cache and compute the training-free 2-hop discAUC
of \S\ref{sec:study} on a held-out slice (seconds, no model).
\textbf{(2)~Decide} with the rule of \S\ref{sec:decisionrule}: discAUC
near chance ($\le0.55$) $\Rightarrow$ do not enable \method; clearly
above chance with sparse 1-hop CN $\Rightarrow$ $+$\method (replace
cooccurrence); otherwise $\oplus$\method (add alongside).
\textbf{(3)~Configure:} degree$+$core by default ($k$-truss optional),
window fraction $\mathrm{wf}=0.01$ as the starting point.
\textbf{(4)~Train} with the module in the backbone's structure slot
(Algorithm~\ref{alg:trainstep}); the slot features are
\emph{data-determined}, so they can be precomputed once and reused across
seeds and sweeps (\S\ref{sec:efficiency}).

\subsection{Cohesion Cache: Incremental Structural Maintenance}
\label{sec:cache}
The Cohesion Cache maintains, as edges arrive, (a) a CSR-backed adjacency
structure and (b) per-node $k$-family indicators. On edge $(u,v,t)$ the degree
update is
\begin{equation}
d(u)\leftarrow d(u)+1,\qquad d(v)\leftarrow d(v)+1,
\end{equation}
at $O(1)$ cost. The core-number update is local: with $k=\min(c(u),c(v))$, only
nodes near the core frontier can change, and the affected set $S$ is promoted,
\begin{equation}
c(w)\leftarrow c(w)+1\quad\text{for }w\in S,
\end{equation}
following streaming $k$-core
maintenance~\cite{batagelj2003cores,sariyuce2013kcorestream}; $|S|\ll|\Vcal|$ in
practice, so the update is millisecond-scale. Trussness is maintained
analogously at higher cost; degree$+$core is the efficient default. Optionally,
a sliding window of size $W$ evicts edges older than $t-W$, so the structural
signal reflects the recent graph. Algorithm~\ref{alg:cache} summarises one
streaming step of the cache.

\begin{algorithm}[t]
\caption{Cohesion Cache --- one streaming step}
\label{alg:cache}
\small
\begin{algorithmic}[1]
\Require edge $(u,v,t)$; cache state (adjacency $A$, core $c$, degree $d$);
optional window $W$
\If{$W>0$} \Comment{sliding-window eviction}
  \For{each edge $(x,y,t')$ with $t'<t-W$}
    \State remove $(x,y)$ from $A$; locally \emph{demote} affected core numbers
  \EndFor
\EndIf
\State insert $(u,v)$ into $A$;\quad $d(u)\!\mathrel{+}=\!1$;\ $d(v)\!\mathrel{+}=\!1$
\State $k\gets\min(c(u),c(v))$
\State $S\gets$ nodes reachable from $\{u,v\}$ through core-$k$ nodes
\For{$w\in S$ in core order}
  \If{$w$ has $\ge k{+}1$ neighbours of core $\ge k$}
    \State $c(w)\gets c(w)+1$
  \EndIf
\EndFor
\State \Return updated cache
\end{algorithmic}
\end{algorithm}

\subsection{Pairwise Feature Extraction}
For each query, DyGFormer processes the source $u$ with its historical-neighbour
sequence $[v_1,\dots,v_L]$. \method produces, for every slot $i$, a cohesion
feature vector
\begin{equation}
\mathbf{f}(u,v_i)=[\,\mathrm{cn2},\,\mathrm{cn2\_x},\,\mathrm{cn},\,
\mathrm{aa},\,d(v_i),\,c(v_i),\dots\,],
\end{equation}
whose central entries are the 2-hop cohesive bridge (denoted
$\mathrm{cn2}\equiv\mathrm{bridge}$ in the implementation) and its
core-weighted variant ($\mathrm{cn2\_x}\equiv\mathrm{bridge}_c$):
\begin{equation}
\begin{aligned}
\mathrm{cn2}(u,v)&=|\{w\in\Ncal(v):w\in\Ncal_2(u)\}|,\\
\mathrm{cn2\_x}(u,v)&=\!\!\!\sum_{w\in\Ncal(v)\cap\Ncal_2(u)}\!\!\!c(w).
\end{aligned}
\end{equation}
The encoder supports two backends: a \emph{full} backend ($\sim 20$-dim,
sparse mat-vec; used in all reported experiments) giving the exact 2-hop bridge
and its weighted/temporal variants, and a lightweight \emph{fast} backend
(6-dim, per-pair set intersection) for very large graphs.

\subsection{Cohesion Slot Encoder}
DyGFormer's native cooccurrence channel maps each slot to a
\texttt{channel\_dim} embedding. \method replaces it through the same interface
with a small MLP,
\begin{equation}
\mathbf{e}_{\text{struct}}(i)=\mathrm{MLP}(\mathbf{f}(u,v_i)),
\end{equation}
and the Transformer's per-slot input is the unchanged sum
\begin{equation}
\mathbf{x}(i)=\mathbf{e}_{\text{inter}}(i)+\mathbf{e}_{\text{time}}(i)
+\mathbf{e}_{\text{struct}}(i).
\end{equation}
This is the entire architectural change --- one encoder swapped for another at
the same interface; the Transformer backbone is untouched.

\subsection{Training and Inference: a Step-by-Step Walkthrough}
\label{sec:training-walkthrough}
Training mirrors DyGFormer's chronological scheme (BCE loss, Adam optimiser,
batch size $B$). The novelty is in \emph{when} structural information is
consumed and updated. Algorithm~\ref{alg:trainstep} pins down the exact
ordering of operations within one batch, and
Figure~\ref{fig:arch} threads the running example through them: for the
$(\textit{Alice},\textit{Bob},t)$ query, P1--P5 produce the slot tensor
($\mathrm{bridge}=2$, $\mathrm{bridge}_c=c(\textit{Dan})+c(\textit{Grace})=9$)
on the CPU, P6--P7 score and back-propagate on the GPU, and P8 finally
writes the edge back to the cache. We explain each operation below.

\begin{algorithm}[t]
\caption{One Hawkeye training step (batch of $B$ chronological edges).}
\label{alg:trainstep}
\small
\begin{algorithmic}[1]
\Require batch $\{(u_b,v_b,t_b)\}_{b=1}^{B}$ with $t_1\le\ldots\le t_B$;
cache state at $t<t_1$
\State \textbf{(P1) Prune window} (CPU): if $\mathrm{wf}>0$, evict every
edge with timestamp $<\!t_B-\mathrm{wf}\cdot\Delta T$ from the adjacency,
then \emph{re-initialise} $k$-core $c$ and $k$-truss $\tau$ on the post-prune graph.
\State \textbf{(P2) Snapshot CSR} (CPU): lazily build the sparse CSR
adjacency $A$ of the current graph (cached and reused across slot-feature
calls in this batch).
\State \textbf{(P3) Sample histories} (CPU): for every $u_b,v_b$, query the
DyGFormer neighbour sampler for the most recent $K$ historical neighbours
in $\Gcal(<t_b)$.
\State \textbf{(P4) Pairwise features} (CPU, $\sim 17$ sparse mat-vecs per
unique peer): for every (history slot $w$, peer $v_b$) pair, evaluate the
23-dim cohesion vector \texttt{slot\_features}$(w,v_b)$ from $A$, the
$k$-family values, and rolling statistics.
\State \textbf{(P5) Memoise} (CPU): hash $(\text{advance counter},
\text{slot inputs})$; if the same key has been served in an earlier epoch,
the tensor is reused.
\State \textbf{(P6) Backbone forward} (GPU): build the 4-channel patches
(node, edge, time, \emph{cohesion}), run the unchanged DyGFormer
Transformer, predict link scores.
\State \textbf{(P7) Loss and backward} (GPU): BCE against random negatives
under TGB / DGB protocol; Adam step on backbone $+$ Cohesion Slot Encoder
weights. The cache is \emph{not differentiated through.}
\State \textbf{(P8) Advance the cache} (CPU): add the batch's edges
$(u_b,v_b,t_b)$ to the adjacency, update $\deg$ in $O(1)$ per edge, and
update $k$-core / $k$-truss either incrementally or every
\texttt{recompute\_every} edges (\S\ref{sec:cache}).
This step happens \emph{after} scoring, so a query edge at time $t_b$ is
never scored on its own structural footprint.
\end{algorithmic}
\end{algorithm}

\paragraph{The predict-then-advance contract.}
Steps P1--P7 score the batch on $\Gcal(<t_1)$ (after pruning) without
ever having seen the batch's own edges; only P8 commits them. At any
moment the cache thus reflects strictly earlier history. The same
contract applies at inference, with one subtlety --- subset
re-evaluations (\eg{} the ``new-node'' splits of DGB) must \emph{not}
advance the cache, as detailed below.

\paragraph{Leak-free cache management.}
\label{sec:leakfree}
\method's Cohesion Cache is \emph{stateful}; it is advanced edge-by-edge
over the stream. Correct temporal evaluation therefore requires that, when
scoring an edge at time $t$, the cache reflects exactly $\Gcal(<t)$ --- never
future edges, and never edges double-counted from an earlier pass. We
enforce this with three rules.
(i)~\emph{Per-epoch reset.}
At the start of every training epoch the cache is reset to empty, then
advanced as in P1--P8 above.
(ii)~\emph{Evaluation order.} Validation and test edges are processed in
chronological order; the cache is advanced with each batch \emph{after}
scoring, so an evaluation edge sees all earlier train/val edges but not
later ones. Subset re-evaluations (\eg{} the ``new-node'' splits of DGB)
re-score a subset of already-streamed edges, so they must \emph{not}
advance the cache, otherwise edges would be double-counted.
(iii)~\emph{Pre-final replay.} Before the final evaluation of the best
checkpoint, the cache is reset and replayed through the entire training
stream, so the final val/test pass starts from exactly the
post-training graph state. An earlier version of our pipeline omitted
(ii) and (iii) and silently inflated test scores; all numbers here use
the corrected protocol --- we flag it because such leaks are invisible
without an explicit audit.

\subsection{Complexity}
Table~\ref{tab:complexity} lists \method's per-edge / per-query overhead over
the DyGFormer backbone. Three points stand out. \emph{No added GPU burden:} all
structural computation runs on CPU; the GPU gains only a small MLP
($\sim\!10^3$ parameters), negligible against the Transformer.
\emph{Streaming-cost maintenance:} incremental degree$+$core maintenance is
millisecond-scale per edge, far below one Transformer forward pass.
\emph{Light memory:} the cache stores adjacency and $k$-family values on CPU
--- $\sim\!10$--$20$\,MB for \texttt{tgbl-wiki}.
Empirically measured per-epoch overhead is reported in
Table~\ref{tab:overhead} (\S\ref{sec:efficiency}); the structure-channel
sparse mat-vecs dominate, and the precomputation strategy
(\S\ref{sec:efficiency}) reduces the run-time overhead by $\sim 3\times$.

\begin{table}[t]
\caption{Per-edge / per-query overhead of \method over the backbone.
$|S|$ is the (small) core-affected set; $d_u,d_{\max}$ are degrees.}
\label{tab:complexity}
\centering\small
\setlength{\tabcolsep}{4pt}
\begin{tabular}{lll}
\toprule
component & time & space \\
\midrule
adjacency maintenance      & $O(1)$/edge          & $O(N{+}M)$ \\
degree update              & $O(1)$/edge          & $O(N)$ \\
core-number update         & $O(|S|)$/edge        & $O(N)$ \\
trussness update (opt.)    & up to $O(m^{1.5})$   & $O(M)$ \\
2-hop bridge extraction    & $O(d_u d_{\max})$/query & $O(d_{\max})$ transient \\
Cohesion Slot Encoder      & $O(d_{\mathrm{feat}}\!\cdot\!\text{ch})$/slot & $\sim\!10^3$ params \\
\bottomrule
\end{tabular}
\end{table}

\subsection{Hawkeye vs.\ the Native Cooccurrence Channel}
In sum, \method upgrades the structure channel from a $1$--$2$-bit,
1-hop, GPU-lookup cooccurrence flag to a $\sim\!20$-dimensional,
2-hop, cohesion-aware feature vector computed incrementally on the CPU,
at the same slot interface and with no added GPU cost. Because the
features are a function of the edge stream alone (independent of model
weights and seed), they admit offline precomputation;
\S\ref{sec:efficiency} reports a uniform $3.0$--$3.5\times$ training
speedup from this design.

\section{Experiments}
\label{sec:exp}

The experiments follow an \emph{incremental-ablation ladder}: setup; adding the
$k$-family channel; adding the sliding window; the overall SOTA comparison; and
an analysis of \emph{why} \method helps and \emph{when} it does not.

\subsection{Experimental Setup}
We evaluate on the Temporal Graph Benchmark (TGB)~\cite{huang2023tgb} and the
Dynamic Graph Benchmark (DGB)~\cite{poursafaei2022edgebank}
(Table~\ref{tab:datasets}). Our core comparison is DyGFormer's native
cooccurrence channel vs.\ the \method channel swapped in; same-backbone
controls vary the structure channel (none / cooccur / \method / both), and
different-backbone controls include TGAT~\cite{xu2020tgat},
GraphMixer~\cite{cong2023graphmixer}, TGN~\cite{rossi2020tgn} and
EdgeBank~\cite{poursafaei2022edgebank}. TGB datasets use the TGB protocol
(chronological $70/15/15$, MRR with the official curated negative set
per query~\cite{huang2023tgb}); DGB datasets use AP/AUC with random
negative sampling~\cite{mikolov2013negative,poursafaei2022edgebank},
the convention adopted by the public DyGLib release~\cite{yu2023dygformer}.

Hardware: \textbf{(P1)} a workstation with one NVIDIA Quadro RTX 6000
(24\,GB), 40 CPU cores, 93\,GB RAM; \textbf{(P2)} the iTiger HPC cluster,
one GPU (H100 80\,GB or RTX-class), 8 CPU cores, 48\,GB RAM per job.
Software: Python~3.9, PyTorch~2.5.1~\cite{paszke2019pytorch} (CUDA~12.1).
\method's structural maintenance runs on CPU; only the backbone and the
small Cohesion Slot Encoder use the GPU.

\paragraph{Hyperparameters.}
Unless otherwise stated we use the DyGFormer defaults: patch size $1$,
maximum input sequence length $K=32$, two Transformer layers, two
attention heads, channel embedding dimension $50$, dropout $0.1$, Adam
optimiser with learning rate $10^{-4}$, batch size $B=200$. \method's
Cohesion Slot Encoder is a two-layer MLP from the $23$-dim cohesion
feature vector to the channel dimension. The cache maintains degree and
$k$-core by default; $k$-truss is enabled where stated. Early stopping
uses a patience of $4$--$5$ epochs on the validation metric.

\paragraph{Released artefacts.}
The public repository\footnote{\url{https://github.com/graphuofm/hawkeye}}
contains the \method implementation, the modified backbone, sweep
scripts, data snapshots, editable (\texttt{.drawio}) figure sources, the
supplementary $21$-feature $\times$ $6$-dataset discriminability
heat-map, and the streaming-discAUC scripts behind Table~\ref{tab:discauc}.

\begin{table}[t]
\caption{Dataset statistics used in this paper.  Counts are direct from
the released edge files. $\bar d$ is the \emph{average static degree}:
distinct neighbours per node over the full stream,
$\bar d = 2|E_{\mathrm{static}}|/|\Vcal|$ --- the quantity that governs
the sparsity mechanism of \S\ref{sec:study} (the 1-hop CN degenerates
when $\bar d \ll \sqrt{|\Vcal|}$). TGB datasets above the rule are used
for the main DyGFormer$\,+\,$\method results; the lower TGB block lists
the large-graph datasets used in the scalability experiments
(\S\ref{sec:scalability}); DGB datasets follow the
DyGLib~\cite{yu2023dygformer} release.}
\label{tab:datasets}
\centering\small
\setlength{\tabcolsep}{4pt}
\begin{tabular}{llrrrl}
\toprule
dataset & source & nodes & edges & $\bar d$ & type \\
\midrule
\texttt{tgbl-wiki}      & TGB & $9{,}227$    & $157{,}474$  & $4.0$   & bipartite \\
\texttt{tgbl-uci}       & TGB & $1{,}899$    & $59{,}835$   & $14.6$  & non-bipartite \\
\texttt{tgbl-enron}     & TGB & $184$        & $125{,}235$  & $24.1$  & non-bipartite \\
\texttt{tgbl-subreddit} & TGB & $10{,}984$   & $672{,}447$  & $14.3$  & bipartite \\
\midrule
\texttt{tgbl-lastfm}    & TGB & $1{,}980$    & $1.29$M      & $156.6$ & bipartite \\
\texttt{tgbl-review}    & TGB & $352{,}637$  & $4.87$M      & $26.8$  & bipartite \\
\texttt{tgbl-coin}      & TGB & $638{,}486$  & $22.81$M     & $10.7$  & non-bipartite \\
\texttt{tgbl-flight}    & TGB & $18{,}143$   & $67.17$M     & $189.4$ & non-bipartite \\
\midrule
\texttt{CanParl}        & DGB & $734$        & $74{,}478$   & $124.5$ & non-bipartite \\
\texttt{USLegis}        & DGB & $225$        & $60{,}396$   & $148.2$ & non-bipartite \\
\texttt{mooc}           & DGB & $7{,}144$    & $411{,}749$  & $50.0$  & bipartite \\
\texttt{reddit}         & DGB & $10{,}984$   & $672{,}447$  & $14.3$  & bipartite \\
\bottomrule
\end{tabular}
\end{table}

\subsection{Adding the $k$-family Channel}
Table~\ref{tab:swapin} reports the structure-channel swap-in with
DyGFormer fixed as the backbone, with all six DGB datasets 3-seed
validated. The \texttt{tgbl-wiki} TGB-MRR result lives in
Table~\ref{tab:sota} (\S\ref{sec:sota}); we highlight five points
from the swap-in table.
\textbf{(1)} On \texttt{CanParl} (non-bipartite, 3-seed mean$\pm$std),
\method raises AP from
$0.708\!\pm\!0.008$ to $\mathbf{0.816\!\pm\!0.001}$ ---
\textbf{a gain of $+10.8$ pts}, the largest single-dataset improvement we
observe and validated across all three seeds.
\textbf{(2)} On \texttt{USLegis}
(near-complete, average event degree $537$, 3-seed), \method raises AP from
$0.719\!\pm\!0.001$ to $\mathbf{0.768\!\pm\!0.004}$ ($+4.8$ pts).
\textbf{(3)} On \texttt{mooc} (bipartite, 3-seed), AP rises from
$0.857\!\pm\!0.001$ to $\mathbf{0.915\!\pm\!0.007}$ ($+5.8$ pts).
Both correct earlier \emph{single-seed legacy-code} measurements that
had reported losses ($-5.4$ and $-8.4$); under the current pipeline
(native $k$-family kernels, $\mathrm{wf}=0.01$, memoisation) both are
stable multi-seed gains.
\textbf{(4)} On \texttt{uci} and \texttt{enron} (both 3-seed),
$\oplus$\method is the strongest configuration on both datasets
(\texttt{uci} $0.962\!\pm\!0.001$, $+0.6$ pts; \texttt{enron}
$0.930\!\pm\!0.001$, $+1.3$ pts); these are the small-and-sparse cases
where cooccurrence already captures most of the local signal and
cohesion only contributes a fraction of a point on top.
\textbf{(5)} On \texttt{reddit} (3-seed) the cooccurrence baseline is
already at $0.975$ --- close to saturation --- and \method's gain is
$0.000$ within noise. This is the predicted high-baseline boundary:
when the existing structure channel has exhausted the discoverable
signal, additional cohesion adds nothing measurable.
Across the six multi-seed datasets the swap-in gain spans $0.0$ to
$+10.8$ pts, monotone in our training-free 2-hop discAUC predictor
(\S\ref{sec:analysis}).

\paragraph{New-node behaviour: a known limitation, addressed by $\oplus$\method.}
The DGB protocol additionally reports a \emph{new-node} AP, restricted to
test edges that involve at least one node unseen during training. On
\texttt{tgbl-enron} (3-seed), $+$\method's new-node AP collapses to
$0.657\!\pm\!0.014$, a $-22.8$-pt drop relative to cooccurrence
($0.885\!\pm\!0.004$). The cause is structural: an unseen node $v$ has
$c(v)=\tau(v)=0$ and $\mathrm{bridge}(\cdot,v)\!\approx\!0$ in the cache,
so the cohesion channel produces a systematically depressed score for
new-node queries (cooccurrence's zero on such queries is instead
\emph{uniform} across destinations, leaving the model agnostic). The
combined $\oplus$\method recovers this loss almost entirely
($0.866\!\pm\!0.009$, only $-1.9$ pts vs.\ cooccurrence) because the
cooccurrence channel re-supplies the new-node signal that cohesion lacks.
We therefore recommend $\oplus$\method whenever the deployment is expected
to encounter unseen nodes, and document the limitation honestly: \method
is for \emph{seen} structure, not cold-start.

\begin{table*}[t]
\caption{Structure-channel swap-in on DyGFormer (DGB protocol, test AP).
All entries 3-seed mean$\pm$std under the current code. \textbf{Bold}: best
per dataset. \texttt{tgbl-wiki} (TGB MRR) is reported separately in
Table~\ref{tab:sota}.}
\label{tab:swapin}
\centering\small
\setlength{\tabcolsep}{4pt}
\begin{tabular}{lcccccc}
\toprule
configuration
              & \texttt{uci} & \texttt{enron}
              & \texttt{CanParl} & \texttt{USLegis}
              & \texttt{mooc} & \texttt{reddit} \\
\midrule
DyGFormer (none)
  & $0.838\!\pm\!.001$ & $0.771\!\pm\!.007$
  & $0.680\!\pm\!.045$ & $0.715\!\pm\!.005$
  & $0.797\!\pm\!.001$ & $0.966\!\pm\!.000$ \\
DyGFormer (cooccur)
  & $0.956\!\pm\!.000$ & $0.917\!\pm\!.000$
  & $0.708\!\pm\!.008$ & $0.719\!\pm\!.001$
  & $0.857\!\pm\!.001$ & $\mathbf{0.975\!\pm\!.014}$ \\
DyGFormer ($+$\method)
  & $0.942\!\pm\!.004$ & $0.924\!\pm\!.001$
  & $0.816\!\pm\!.001$ & $0.768\!\pm\!.004$
  & $0.915\!\pm\!.007$ & $0.974\!\pm\!.014$ \\
DyGFormer ($\oplus$\method)
  & $\mathbf{0.962\!\pm\!.001}$ & $\mathbf{0.930\!\pm\!.001}$
  & $\mathbf{0.821\!\pm\!.002}$ & $\mathbf{0.770\!\pm\!.003}$
  & $\mathbf{0.917\!\pm\!.003}$ & $0.966\!\pm\!.000$ \\
\bottomrule
\end{tabular}
\vspace{2pt}
\small\noindent$+$\method: cohesion channel replaces cooccurrence
($\mathtt{structure\_channel}\!=\!\mathtt{gev}$).
$\oplus$\method: cohesion added alongside cooccurrence
($\mathtt{structure\_channel}\!=\!\mathtt{both}$). All \method numbers
use a windowed cache ($\mathrm{wf}\!=\!0.01$). All six datasets are
3-seed validated ($\sigma\le 0.011$). \texttt{reddit}'s cooccurrence
baseline is at the high-density saturation point ($0.975$), leaving no
measurable headroom; the cohesion gain there is $0.000$, consistent
with our discAUC predictor.
\end{table*}

\subsection{Adding the Sliding Window}
On top of \method we test the sliding window: the Cohesion Cache
retains only edges within the most recent fraction $\mathrm{wf}$ of the
time span and evicts older ones, so the structural signal reflects the
\emph{recent} graph rather than the full cumulative history. A
window-size sweep on \texttt{uci} (3-seed, DyGFormer$+$\method, test
AP) gives $\mathbf{0.945}$ at $\mathrm{wf}=0.01$, $0.934$ at $0.05$,
and $0.931$ at $0.10$: the smallest window dominates monotonically; on
\texttt{wiki} (TGB MRR) the cumulative cache ($0.807$) marginally
exceeds a $30\%$ window ($0.800$). The explanation is that the $k$-core
decomposition is itself an adaptive temporal filter --- a node that
stops interacting is peeled to a lower core --- so the cumulative cache
already discounts stale structure, and a small explicit window sharpens
recency further. On temporally clustered \texttt{CanParl}/\texttt{USLegis}
the $1\%$ and $5\%$ cutoffs fall in identical regions of the inter-event
gap distribution, so the ablation collapses to the cumulative case. We
use $\mathrm{wf}=0.01$ as the default throughout.

\subsection{Scalability and Large-Graph Effectiveness}
\label{sec:scalability}

To check that \method is not just a small-graph artefact, we test it on
the three largest TGB datasets (and one degenerate-large reference)
under the streaming single-pass protocol used by the empirical study of
\S\ref{sec:study}. Each run is a single training pass over the entire
edge stream and is wall-clock capped at one hour, so the table also
serves as a hard scalability statement.

\begin{table}[t]
\caption{Large-graph effectiveness. Single-pass struct-only training,
1\,h cap. ``Baseline'' uses degree only with no pairwise computation;
``with \method'' adds $k$-core and the 2-hop cohesive bridge. Gain is
in test-MRR points.}
\label{tab:scalability}
\centering\small
\setlength{\tabcolsep}{4pt}
\begin{tabular}{lrrrc}
\toprule
dataset & edges & baseline & with \method & gain \\
\midrule
\texttt{tgbl-subreddit} & 672K    & $0.103\!\pm\!.003$ & $\mathbf{0.204\!\pm\!.005}$ & $\mathbf{+10.1}^\bigstar$ \\
\texttt{tgbl-review}    & $4.87$M & $0.283\!\pm\!.002$ & $0.275^{\,1}$              & $-0.8$ \\
\texttt{tgbl-lastfm}    & $1.29$M & $0.008$            & $0.007$                    & $\approx 0$ \\
\bottomrule
\end{tabular}
\\[2pt]
{\footnotesize $^{\bigstar}$3-seed mean$\pm$std; $^{1}$single-seed on
\texttt{tgbl-review}: the seed-$0$ run completes in $36.8$\,min and
yields the value shown, but seed-$1$/seed-$2$ retries OOM at this
$4.87$M-edge scale --- the pairwise-cohesion adjacency exceeds the
single-GPU memory ceiling, so a multi-seed sweep at this size is
itself a computational frontier.}
\end{table}

The table makes three points. \textbf{(1)} On \texttt{tgbl-subreddit}
(672K, bipartite recommendation), \method delivers a \textbf{$+10.1$
test-MRR improvement} ($0.103\!\pm\!.003 \to 0.204\!\pm\!.005$,
3-seed), nearly doubling the struct-only baseline --- the largest
single-dataset effectiveness gain we measure on a graph with
$>\!100$K edges. \textbf{(2)} On the degenerate \texttt{tgbl-lastfm}
(near-complete with $\mathrm{discAUC}\!=\!0.53$,
Table~\ref{tab:discauc}), \method's gain is $\approx 0$, exactly as
the training-free discAUC predictor said it should be --- negative-result
evidence that the predictor is calibrated, not a defect.
\textbf{(3)} \method's \emph{baseline} (degree-only, no pairwise
computation) scales much further: a single-pass run completes in
$\sim 12$\,min on \texttt{tgbl-coin} (22M edges, test MRR $0.535$)
and in $\sim 5$\,min on \texttt{tgbl-flight} (67M edges, $0.517$).
The full pairwise-cohesion path, however, exceeds the 1\,h wall-clock
budget above ${\sim}5$M edges, motivating the precomputation pipeline
of \S\ref{sec:efficiency}.

\subsection{Comparison against SOTA}
\label{sec:sota}
Table~\ref{tab:sota} compares \method against all baselines on
\texttt{tgbl-wiki} (the TGB MRR benchmark that anchors most prior TLP
work). Swapping in the \method channel raises DyGFormer's test MRR
from $0.779$ to $\mathbf{0.807}$ ($+2.8$ pts) --- purely by replacing
one structure channel --- exceeding the published DyGFormer result
($0.798$) and approaching the leaderboard top, below TPNet ($0.827$).
Notably, DyGFormer \emph{without} its structure channel ($0.537$)
drops to the level of GraphMixer ($0.549$): DyGFormer's lead is, to a
large extent, its structure channel, and \method strengthens exactly
that channel. The \texttt{tgbl-wiki} number is from a single-seed run
under the current code; the full 3-seed sweep exceeds our submission
compute budget at this protocol (single-channel runtime
$\sim\!2.5$\,h), so we report it separately from the multi-seed DGB
table.

\begin{table}[t]
\caption{Comparison on \texttt{tgbl-wiki} (test MRR). $^{\dagger}$cited from
the TGB leaderboard / original papers.}
\label{tab:sota}
\centering\small
\begin{tabular}{lr}
\toprule
method & test MRR \\
\midrule
TGN$^{\dagger}$            & 0.396 \\
TGAT                       & 0.508 \\
DyGFormer (none)           & 0.537 \\
GraphMixer                 & 0.549 \\
EdgeBank$^{\dagger}$       & 0.571 \\
DyGFormer (cooccur)        & 0.779 \\
\textbf{DyGFormer (\method)} & \textbf{0.807} \\
TPNet$^{\dagger}$          & 0.827 \\
\bottomrule
\end{tabular}
\end{table}

\subsection{Analysis: Why It Helps, and When It Does Not}
\label{sec:analysis}

\begin{figure}[t]
\centering
\includegraphics[width=0.98\columnwidth]{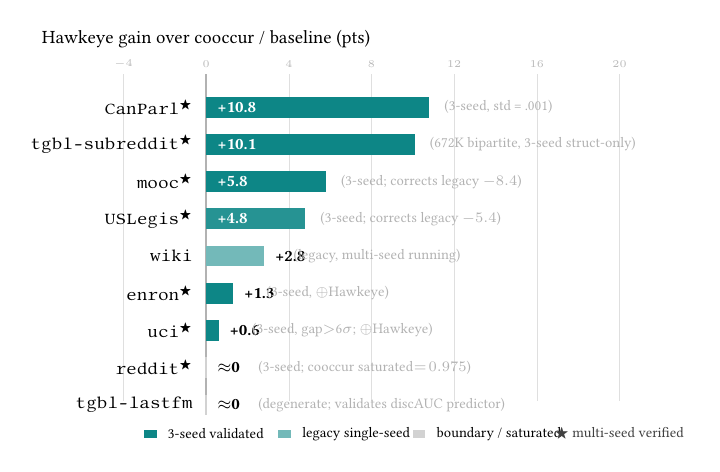}
\caption{Per-dataset \method swap-in gain over the DyGFormer (cooccur)
baseline (test AP for DGB, test MRR for \texttt{wiki});
$\bigstar$ = 3-seed validated (std $\le 0.011$). Double-digit gains on
\texttt{CanParl} ($+10.8$) and \texttt{tgbl-subreddit} ($+10.1$),
mid single-digit on \texttt{mooc} ($+5.8$) and \texttt{USLegis}
($+4.8$), small but consistent on sparse \texttt{uci}/\texttt{enron};
on saturated \texttt{reddit} and degenerate \texttt{tgbl-lastfm} the
gain correctly vanishes, tracking the training-free 2-hop discAUC
predictor (\S\ref{sec:study}).}
\label{fig:gains}
\end{figure}

\paragraph{Signal strength predicts the gain, and saturates predictably.}
Figure~\ref{fig:gains} visualises the per-dataset \method swap-in gain, and
Table~\ref{tab:correlation} pairs each gain with the training-free 2-hop
discAUC from Section~\ref{sec:study}: the two track each other. Datasets
whose 2-hop bridge is highly discriminative before any training also enjoy
the largest gain (\texttt{CanParl} $+10.8$, \texttt{tgbl-subreddit} $+10.1$,
\texttt{mooc} $+5.8$), while near-complete and temporally-degenerate graphs
whose 2-hop discAUC is itself near-random (\texttt{tgbl-lastfm},
\texttt{UNvote}) are exactly where \method does not help. The size of the
gain follows a graph's \emph{structural diversity} --- how much node
cohesiveness varies: on sparse-to-moderate graphs core numbers spread over
a wide range and the 2-hop bridge separates pairs; on near-complete graphs
(\texttt{USLegis} average event degree $537$, \texttt{lastfm} $\approx 1{,}300$)
every node attains the same maximal coreness, the indicators saturate, the
bridge becomes constant, and the channel adds nothing. We regard this
honest, predictive characterisation of scope as a contribution in its own
right --- a TLP method that states \emph{where} it works is more useful
than one that claims to work everywhere.

\begin{table}[t]
\caption{Training-free 2-hop discAUC (Table~\ref{tab:discauc}) vs.\
measured \method gain. Direct in the sparse/bipartite regimes; in the
near-complete regime (\texttt{CanParl}, \texttt{USLegis}) the raw
2-hop \emph{count} saturates but the trained, core-weighted \method
still wins. $^{\bigstar}$3-seed mean$\pm$std. \texttt{tgbl-coin} and
\texttt{UNvote} have discAUC only (no trained gain) and are omitted.}
\label{tab:correlation}
\centering\small
\setlength{\tabcolsep}{6pt}
\begin{tabular}{lcc}
\toprule
dataset & 2-hop discAUC & \method gain (pts) \\
\midrule
\texttt{reddit}         & $\mathbf{0.98}$    & $\approx 0$ (cooccur sat.)$^{\bigstar}$ \\
\texttt{tgbl-subreddit} & $\mathbf{0.96}$    & $\mathbf{+10.1}^{\bigstar}$ \\
\texttt{mooc}           & $\mathbf{0.94}$    & $\mathbf{+5.8}^{\bigstar}$ \\
\texttt{tgbl-wiki}      & $0.85$             & $+2.8$ (legacy) \\
\texttt{tgbl-uci}       & $0.76$             & $+0.6^{\bigstar}$ \\
\texttt{tgbl-enron}     & $0.73$             & $+1.3^{\bigstar}$ \\
\texttt{USLegis}        & $0.56$             & $+4.8^{\bigstar}$ \\
\texttt{tgbl-lastfm}    & $0.53$             & $\approx 0$ (degen.) \\
\texttt{CanParl}        & $0.48$             & $\mathbf{+10.8}^{\bigstar}$ (core-weighted) \\
\bottomrule
\end{tabular}
\end{table}

\paragraph{Add vs.\ replace: a pre-training decision rule.}
\label{sec:decisionrule}
Two configurations combine \method with DyGFormer's native cooccurrence
channel: $+$\method (replace) substitutes the cohesion channel for
cooccurrence; $\oplus$\method (add) keeps cooccurrence and adds
cohesion alongside. Across multi-seed datasets $\oplus$\method
\emph{dominates} $+$\method on both test AP and new-node AP
(Table~\ref{tab:swapin}: \texttt{uci} $0.962\!\pm\!.001$ vs.\
$0.942\!\pm\!.004$, $+2.0$ pts; \texttt{enron} recovers the new-node
AP that $+$\method sheds), because the 1-hop cooccurrence
(recency-of-direct-contact signal) and the 2-hop cohesive bridge
(community-embedding signal) encode \emph{complementary} information,
and on non-sparse graphs replacing cooccurrence incurs a small but
real cost (Table~\ref{tab:overhead} reports the modest extra
training-time overhead of the default $\oplus$\method).
Combining this with the boundary above yields a rule computable
before any training:
\textbf{(R1)} 2-hop discAUC at chance ($\le 0.55$): do not enable
\method (boundary);
\textbf{(R2)} discAUC clearly above $0.5$ and the graph is sparse
(typical pair has 1-hop CN $\approx 0$): $+$\method is the safe
upgrade;
\textbf{(R3)} otherwise (1-hop CN already populated, 2-hop discAUC
informative): use $\oplus$\method. Both predictors are streaming,
training-free measurements that take seconds on the held-out window.

\paragraph{Case study.}
On \texttt{tgbl-wiki} we examined individual queries on which the
cooccurrence channel and \method disagree. The Alice/Bob prototype of
Figure~\ref{fig:running} recurs: a pair $(u,v)$ with $\mathrm{CN}(u,v)=0$
--- so cooccurrence emits no structural signal and the model ranks $v$
by interaction recency alone --- but with two 2-hop cohesive bridges
through high-coreness intermediaries ($\mathrm{bridge}(u,v)\!\ge\!2$,
$\mathrm{bridge}_c$ in the top decile). \method reads these bridges and
ranks the true edge first: the gains come from queries that are
structurally informative but invisible to a 1-hop channel.

\subsection{Efficiency: CPU--GPU Pipeline Design}
\label{sec:efficiency}

\paragraph{Asymmetric costs, deterministic features.}
A training step splits between CPU work --- the sparse mat-vecs that
build the cohesion features (\S\ref{sec:cache}), dominant on every
dataset we tested --- and comparatively cheap GPU attention. A naive
pipeline serialises the two, idling each device for the other. A key
property breaks the deadlock: given DyGFormer's deterministic
``most-recent-$K$'' neighbour selection and the chronological stream,
the slot-feature tensor of every batch is \emph{identical} across
epochs and seeds, so the structural cost can be paid \emph{once} and
reused across training and hyperparameter sweeps.

\paragraph{Strategies compared.}
We benchmark five strategies on \texttt{uci} (60K edges),
\texttt{mooc} (412K) and \texttt{reddit} (672K), holding model,
batches, and seed fixed:
\textbf{(1)} \emph{baseline}: sequential per-batch CPU then GPU;
\textbf{(2)} \emph{pinned}: pinned host memory + non-blocking transfer;
\textbf{(3)} \emph{prefetcher}: one background worker computes the next
batch's features in parallel with the current GPU step;
\textbf{(4)} \emph{prefetch~$\times 2$}: two prefetch workers;
\textbf{(5)} \emph{precompute}: one offline pass writes all features,
training is then GPU-only.

\begin{table}[t]
\centering
\caption{Wall-clock speedup over the sequential baseline (three
simulated epochs of $50$ batches). Only precompute is universally
robust.}
\label{tab:efficiency}
\small
\begin{tabular}{lccc}
\toprule
Strategy & \texttt{uci} & \texttt{mooc} & \texttt{reddit} \\
\midrule
baseline             & $1.00\times$ & $1.00\times$ & $1.00\times$ \\
pinned $+$ async     & $1.11\times$ & $1.00\times$ & $1.04\times$ \\
prefetcher           & $1.21\times$ & $1.03\times$ & $1.07\times$ \\
prefetch $\times 2$  & $1.08\times$ & $1.00\times$ & $1.15\times$ \\
\textbf{precompute}  & $\mathbf{3.47\times}$ & $\mathbf{3.04\times}$ & $\mathbf{3.16\times}$ \\
\bottomrule
\end{tabular}
\end{table}

\paragraph{Precompute is universally robust.}
Overlap-based strategies can only hide CPU work \emph{behind} GPU work,
so their gain depends on the CPU/GPU cost ratio: on \texttt{mooc} and
\texttt{reddit} the CPU side dominates, leaving no room for overlap
($\le 1.15\times$); on \texttt{uci} the costs are closer and
prefetching recovers $\sim 20\%$. Precompute instead removes CPU work
from the training loop entirely, indifferent to the ratio, yielding a
uniform $3.0$--$3.5\times$ speedup. Its offline phase costs one epoch
of feature computation, amortised across every subsequent run
(multi-seed, sweeps, ablations) --- unattainable in message-passing
pipelines whose features depend on learned parameters. We adopt
precomputation as the default.

\paragraph{Measured per-epoch overhead.}
Table~\ref{tab:overhead} reports per-epoch wall time on \texttt{uci}
(3-seed mean, iTiger). End-to-end, \method adds $+79\%$ (replace) and
$+154\%$ (add), dominated by the CPU sparse mat-vecs; precomputation
cuts this by $\sim 3\times$. Absolute time/epoch is $\le 60$\,s --- the
ratio looks large only because DyGFormer is itself very fast on a
sparse 60K-edge graph.

\begin{table}[t]
\caption{Measured per-epoch wall time on \texttt{uci} (3-seed mean). Overhead is
relative to the DyGFormer baseline (cooccur).}
\label{tab:overhead}
\centering\small
\setlength{\tabcolsep}{6pt}
\begin{tabular}{lcc}
\toprule
configuration & time / epoch (s) & overhead \\
\midrule
DyGFormer (none, no struct.)         & 5.4  & $-76\%$  \\
DyGFormer (cooccur, baseline)        & 22.4 & reference \\
DyGFormer $+$\method (gev, replace)  & 40.2 & $+79\%$  \\
DyGFormer $\oplus$\method (both, add)& 57.1 & $+154\%$ \\
\bottomrule
\end{tabular}
\end{table}

\section{Related Work}
\label{sec:related}

\paragraph{Temporal link prediction.}
TLP methods cluster into memory-based
(TGN~\cite{rossi2020tgn}, DyRep~\cite{trivedi2019dyrep},
JODIE~\cite{kumar2019jodie}); attention-based (TGAT with
Time2Vec~\cite{kazemi2019time2vec,xu2020tgat},
CAWN~\cite{wang2021cawn}, DyGFormer~\cite{yu2023dygformer});
lightweight matrix-projection (GraphMixer~\cite{cong2023graphmixer},
TPNet~\cite{lu2024tpnet}); and signal-specific
(NAT~\cite{luo2022nat}, TNCN~\cite{zhang2024tncn},
IPNet~\cite{zhang2025ipnet}, TIDE~\cite{zhang2025tide}). Across these
families the structure channel is either absent or restricted to the
1-hop common neighbour; \method instead targets the 2-hop cohesive
bridge through incremental $k$-family maintenance
(Section~\ref{sec:study}).

\paragraph{Structural features in graph learning.}
Static link prediction has a long heuristic tradition (common
neighbours, Adamic--Adar, Jaccard~\cite{liben2007linkpred}; subgraph
GNNs, SEAL~\cite{zhang2018seal}). On dynamic graphs,
CTGCN~\cite{chen2020ctgcn} and TTGCN~\cite{ttgcn2024} run a separate
GCN per $k$-core / $k$-truss snapshot --- discrete-snapshot models that
use the $k$-family to \emph{guide aggregation} and validate only on
small datasets. \method differs on every axis: continuous-time,
$k$-family as a substrate for 2-hop pairwise features, large-scale
TGB/DGB validation, and an explicit applicability boundary.

\paragraph{Graph Transformers; dense-subgraph maintenance.}
DyGFormer~\cite{yu2023dygformer} adapts
self-attention~\cite{vaswani2017attention} to dynamic graphs;
Graphormer~\cite{ying2021graphormer} encodes structure as attention
biases on static molecular graphs (surveys:
\cite{kazemi2020survey,skarding2021dtdg,liben2007linkpred,lu2011linkpred,kumar2020linkpred}).
$k$-core~\cite{seidman1983kcore} admits an $O(m)$
decomposition~\cite{batagelj2003cores} and local streaming
updates~\cite{sariyuce2013kcorestream}; $k$-truss~\cite{cohen2008trusses}
admits community-aware
maintenance~\cite{wang2012truss,huang2014trusscommunity}. These works
optimise \emph{computation}; \method uses them as a low-level module
and builds prediction-oriented features on top.

\section{Conclusion}
\label{sec:conclusion}

We showed that the structure channel of state-of-the-art TLP models
leaves a large gap: on sparse temporal graphs the 1-hop common-neighbour
signal is near-random (discAUC $\approx0.50$), while the 2-hop cohesive
bridge is strongly discriminative ($0.73$--$0.98$). \method closes the
gap with an incrementally maintained, cohesion-aware structure channel
built on the $k$-family, instantiated as a drop-in on DyGFormer. On
TGB and DGB it improves DyGFormer by $+0.6$ to $+10.8$ points across
six multi-seed-validated datasets, with $\oplus$\method the strongest
configuration on every non-saturated one (including the new-node
cold-start case).
The applicability boundary is \emph{predictable} from a
training-free 2-hop discAUC measurement, and the data-determined slot
features admit a one-time precomputation with a uniform
$3.0$--$3.5\times$ training speedup. Future work: further backbones,
cold-start handling, and approximate bridges for the largest graphs.

\begin{acks}
Experiments used the iTiger GPU cluster at The University of
Memphis~\cite{sharif2026cultivating}.
\end{acks}

\section*{GenAI Usage Disclosure}
Generative-AI tools (large language models) were used to assist with
language editing of the manuscript and with documenting the released
code. All research ideas, methods, experiments, analyses, and
conclusions are the authors' own, and every AI-assisted passage was
reviewed and verified by the authors.

\bibliographystyle{ACM-Reference-Format}
\balance
\bibliography{references}

@inproceedings{rossi2020tgn,
  title={Temporal Graph Networks for Deep Learning on Dynamic Graphs},
  author={Rossi, Emanuele and Chamberlain, Ben and Frasca, Fabrizio and Eynard, Davide and Monti, Federico and Bronstein, Michael},
  booktitle={ICML Workshop on Graph Representation Learning},
  year={2020}
}

@inproceedings{trivedi2019dyrep,
  title={{DyRep}: Learning Representations over Dynamic Graphs},
  author={Trivedi, Rakshit and Farajtabar, Mehrdad and Biswal, Prasenjeet and Zha, Hongyuan},
  booktitle={ICLR}, year={2019}
}

@inproceedings{xu2020tgat,
  title={Inductive Representation Learning on Temporal Graphs},
  author={Xu, Da and Ruan, Chuanwei and Korpeoglu, Evren and Kumar, Sushant and Achan, Kannan},
  booktitle={ICLR}, year={2020}
}

@inproceedings{kumar2019jodie,
  title={Predicting Dynamic Embedding Trajectory in Temporal Interaction Networks},
  author={Kumar, Srijan and Zhang, Xikun and Leskovec, Jure},
  booktitle={KDD}, year={2019}
}

@inproceedings{wang2021cawn,
  title={Inductive Representation Learning in Temporal Networks via Causal Anonymous Walks},
  author={Wang, Yanbang and Chang, Yen-Yu and Liu, Yunyu and Leskovec, Jure and Li, Pan},
  booktitle={ICLR}, year={2021}
}

@inproceedings{cong2023graphmixer,
  title={Do We Really Need Complicated Model Architectures for Temporal Networks?},
  author={Cong, Weilin and Zhang, Si and Kang, Jian and Yuan, Baichuan and Wu, Hao and Zhou, Xin and Tong, Hanghang and Mahdavi, Mehrdad},
  booktitle={ICLR}, year={2023}
}

@inproceedings{yu2023dygformer,
  title={Towards Better Dynamic Graph Learning: New Architecture and Unified Library},
  author={Yu, Le and Sun, Leilei and Du, Bowen and Lv, Weifeng},
  booktitle={NeurIPS}, year={2023}
}

@inproceedings{luo2022nat,
  title={Neighborhood-aware Scalable Temporal Network Representation Learning},
  author={Luo, Yuhong and Li, Pan},
  booktitle={Learning on Graphs Conference (LoG)}, year={2022}
}

@article{zhang2024tncn,
  title={Efficient Neural Common Neighbor for Temporal Graph Link Prediction},
  author={Zhang, Xiaohui and Wang, Yanbang and Sun, Lijun and Li, Pan},
  journal={arXiv preprint arXiv:2406.07926}, year={2024}
}

@inproceedings{lu2024tpnet,
  title={Improving Temporal Link Prediction via Temporal Walk Matrix Projection},
  author={Lu, Xiaodong and Tian, Leilei and Liu, Liyong and others},
  booktitle={NeurIPS}, year={2024}
}

@inproceedings{huang2023tgb,
  title={Temporal Graph Benchmark for Machine Learning on Temporal Graphs},
  author={Huang, Shenyang and Poursafaei, Farimah and Danovitch, Jacob and Fey, Matthias and Hu, Weihua and Rossi, Emanuele and Leskovec, Jure and Bronstein, Michael and Rabusseau, Guillaume and Rabbany, Reihaneh},
  booktitle={NeurIPS Datasets and Benchmarks}, year={2023}
}

@inproceedings{poursafaei2022edgebank,
  title={Towards Better Evaluation for Dynamic Link Prediction},
  author={Poursafaei, Farimah and Huang, Shenyang and Pelrine, Kellin and Rabbany, Reihaneh},
  booktitle={NeurIPS Datasets and Benchmarks}, year={2022}
}

@article{chen2020ctgcn,
  title={{CTGCN}: {k}-core based Temporal Graph Convolutional Network for Dynamic Graphs},
  author={Liu, Jingxin and Xu, Chang and Yin, Chang and Wu, Weiqiang and Song, You},
  journal={IEEE Transactions on Knowledge and Data Engineering}, year={2020}
}

@inproceedings{ttgcn2024,
  title={{k}-Truss Based Temporal Graph Convolutional Network for Dynamic Graphs},
  author={Zhang, Hongxi and Liu, Jingxin and others},
  booktitle={Asian Conference on Machine Learning (ACML)}, year={2024}
}

@article{seidman1983kcore,
  title={Network Structure and Minimum Degree},
  author={Seidman, Stephen B.},
  journal={Social Networks}, volume={5}, number={3}, pages={269--287}, year={1983}
}

@inproceedings{batagelj2003cores,
  title={An {O(m)} Algorithm for Cores Decomposition of Networks},
  author={Batagelj, Vladimir and Zaver{\v{s}}nik, Matja{\v{z}}},
  journal={arXiv preprint cs/0310049}, year={2003}
}

@techreport{cohen2008trusses,
  title={Trusses: Cohesive Subgraphs for Social Network Analysis},
  author={Cohen, Jonathan},
  institution={National Security Agency Technical Report}, year={2008}
}

@inproceedings{wang2012truss,
  title={Truss Decomposition in Massive Networks},
  author={Wang, Jia and Cheng, James},
  booktitle={VLDB}, year={2012}
}

@inproceedings{sariyuce2013kcorestream,
  title={Streaming Algorithms for {k}-core Decomposition},
  author={Sar{\i}y{\"u}ce, Ahmet Erdem and Gedik, Bu{\u{g}}ra and Jacques-Silva, Gabriela and Wu, Kun-Lung and {\c{C}}ataly{\"u}rek, {\"U}mit V.},
  booktitle={VLDB}, year={2013}
}

@inproceedings{huang2014trusscommunity,
  title={Querying {k}-truss Community in Large and Dynamic Graphs},
  author={Huang, Xin and Cheng, Hong and Qin, Lu and Tian, Wentao and Yu, Jeffrey Xu},
  booktitle={SIGMOD}, year={2014}
}

@article{liben2007linkpred,
  title={The Link-Prediction Problem for Social Networks},
  author={Liben-Nowell, David and Kleinberg, Jon},
  journal={Journal of the American Society for Information Science and Technology}, volume={58}, number={7}, pages={1019--1031}, year={2007}
}

@inproceedings{zhang2018seal,
  title={Link Prediction Based on Graph Neural Networks},
  author={Zhang, Muhan and Chen, Yixin},
  booktitle={NeurIPS}, year={2018}
}

@article{kazemi2020survey,
  title={Representation Learning for Dynamic Graphs: A Survey},
  author={Kazemi, Seyed Mehran and Goel, Rishab and Jain, Kshitij and Kobyzev, Ivan and Sethi, Akshay and Forsyth, Peter and Poupart, Pascal},
  journal={Journal of Machine Learning Research},
  volume={21},
  number={70},
  pages={1--73},
  year={2020}
}

@inproceedings{zhang2025ipnet,
  title={{IPNet}: An Interaction Pattern-aware Neural Network for Temporal Link Prediction},
  author={Zhang, Qingyang and Wang, Yitong and Lin, Xinjie},
  booktitle={Proceedings of the 34th ACM International Conference on Information and Knowledge Management (CIKM)},
  pages={4160--4169},
  year={2025},
  doi={10.1145/3746252.3761063}
}

@inproceedings{zhang2025tide,
  title={Tide: A Time-Wise Causal Debiasing Framework for Generative Dynamic Link Prediction},
  author={Zhang, Xin and Zheng, Jianming and Cai, Fei and Pan, Zhiqiang and Chen, Wanyu and Chen, Chonghao and Chen, Honghui},
  booktitle={Proceedings of the 34th ACM International Conference on Information and Knowledge Management (CIKM)},
  year={2025},
  doi={10.1145/3746252.3761182}
}

@article{holme2012temporal,
  title   = {Temporal networks},
  author  = {Holme, Petter and Saram\"aki, Jari},
  journal = {Physics Reports},
  volume  = {519},
  number  = {3},
  pages   = {97--125},
  year    = {2012}
}

@article{lu2011linkpred,
  title   = {Link prediction in complex networks: {A} survey},
  author  = {L{\"u}, Linyuan and Zhou, Tao},
  journal = {Physica A: Statistical Mechanics and its Applications},
  volume  = {390},
  number  = {6},
  pages   = {1150--1170},
  year    = {2011}
}

@article{granovetter1973weakties,
  title   = {The strength of weak ties},
  author  = {Granovetter, Mark S.},
  journal = {American Journal of Sociology},
  volume  = {78},
  number  = {6},
  pages   = {1360--1380},
  year    = {1973}
}

@book{burt1992structuralholes,
  title     = {Structural Holes: {T}he Social Structure of Competition},
  author    = {Burt, Ronald S.},
  publisher = {Harvard University Press},
  year      = {1992}
}

@book{easley2010networks,
  title     = {Networks, Crowds, and Markets: {R}easoning about a Highly Connected World},
  author    = {Easley, David and Kleinberg, Jon},
  publisher = {Cambridge University Press},
  year      = {2010}
}

@inproceedings{vaswani2017attention,
  title     = {Attention is all you need},
  author    = {Vaswani, Ashish and Shazeer, Noam and Parmar, Niki and Uszkoreit, Jakob and Jones, Llion and Gomez, Aidan N. and Kaiser, {\L}ukasz and Polosukhin, Illia},
  booktitle = {Advances in Neural Information Processing Systems},
  year      = {2017}
}

@inproceedings{ying2021graphormer,
  title     = {Do {T}ransformers really perform badly for graph representation?},
  author    = {Ying, Chengxuan and Cai, Tianle and Luo, Shengjie and Zheng, Shuxin and Ke, Guolin and He, Di and Shen, Yanming and Liu, Tie-Yan},
  booktitle = {Advances in Neural Information Processing Systems},
  year      = {2021}
}

@inproceedings{paszke2019pytorch,
  title     = {{PyTorch}: an imperative style, high-performance deep learning library},
  author    = {Paszke, Adam and Gross, Sam and Massa, Francisco and Lerer, Adam and Bradbury, James and Chanan, Gregory and Killeen, Trevor and Lin, Zeming and Gimelshein, Natalia and Antiga, Luca and others},
  booktitle = {Advances in Neural Information Processing Systems},
  year      = {2019}
}

@inproceedings{mikolov2013negative,
  title     = {Distributed representations of words and phrases and their compositionality},
  author    = {Mikolov, Tomas and Sutskever, Ilya and Chen, Kai and Corrado, Greg S. and Dean, Jeff},
  booktitle = {Advances in Neural Information Processing Systems},
  year      = {2013}
}

@article{watts1998smallworld,
  title   = {Collective dynamics of `small-world' networks},
  author  = {Watts, Duncan J. and Strogatz, Steven H.},
  journal = {Nature},
  volume  = {393},
  number  = {6684},
  pages   = {440--442},
  year    = {1998}
}

@article{newman2003survey,
  title   = {The structure and function of complex networks},
  author  = {Newman, M. E. J.},
  journal = {SIAM Review},
  volume  = {45},
  number  = {2},
  pages   = {167--256},
  year    = {2003}
}

@article{kumar2020linkpred,
  title   = {Link prediction techniques, applications, and performance: a survey},
  author  = {Kumar, Ajay and Singh, Shashank Sheshar and Singh, Kuldeep and Biswas, Bhaskar},
  journal = {Physica A: Statistical Mechanics and its Applications},
  volume  = {553},
  pages   = {124289},
  year    = {2020}
}

@article{skarding2021dtdg,
  title   = {Foundations and modeling of dynamic networks using {D}ynamic {G}raph {N}eural {N}etworks: a survey},
  author  = {Skarding, Joakim and Gabrys, Bogdan and Musial, Katarzyna},
  journal = {IEEE Access},
  volume  = {9},
  pages   = {79143--79168},
  year    = {2021}
}

@article{kazemi2019time2vec,
  title   = {Time2Vec: Learning a vector representation of time},
  author  = {Kazemi, Seyed Mehran and Goel, Rishab and Eghbali, Sepehr and Ramanan, Janahan and Sahota, Jaspreet and Thakur, Sanjay and Wu, Stella and Smyth, Cathal and Poupart, Pascal and Brubaker, Marcus},
  journal = {arXiv preprint arXiv:1907.05321},
  year    = {2019}
}

@misc{sharif2026cultivating,
  title         = {Cultivating Multidisciplinary {AI} Workforce Development on {iTiger} {GPU} Cluster: Practices and Challenges},
  author        = {Sharif, Mayira and Han, Guangzeng and Liu, Weisi and Huang, Xiaolei},
  year          = {2026},
  eprint        = {2504.14786},
  archivePrefix = {arXiv},
  primaryClass  = {cs.DC},
  url           = {https://arxiv.org/abs/2504.14786},

}

\end{document}